\documentclass[5p]{elsarticle}
\pdfoutput=1

\usepackage{lineno}
\PassOptionsToPackage{hyphens}{url}\usepackage[breaklinks]{hyperref}
\modulolinenumbers[5]

\makeatletter
\def\ps@pprintTitle{%
 \let\@oddhead\@empty
 \let\@evenhead\@empty
 \def\@oddfoot{}%
 \let\@evenfoot\@oddfoot}
\makeatother

\usepackage{subcaption}
\usepackage{amsmath}
\usepackage{seqsplit}
\newcommand{\hash}[1]{{\ttfamily\seqsplit{#1}}}
\usepackage{xcolor}
\usepackage[linesnumbered, ruled, vlined]{algorithm2e}
\SetKwRepeat{Do}{do}{while}%
\usepackage{graphicx}
\usepackage{amsmath}
\usepackage{amssymb}
\graphicspath{{imagenes/}}
\usepackage{longtable}
\usepackage{textcomp}
\usepackage[T1]{fontenc}

\newif\ifstatus
\statusfalse

\begin{document}

\begin{frontmatter}

\title{Avaddon ransomware: an in-depth analysis and decryption of infected systems}


\author[URJC]{Javier Yuste\corref{mycorrespondingauthor}}\cortext[mycorrespondingauthor]{Corresponding author}
\ead{javier.yuste@urjc.es}

\author[UC3M]{Sergio Pastrana}

\address[URJC]{Universidad Rey Juan Carlos, Madrid}
\address[UC3M]{Universidad Carlos III, Madrid}

\begin{abstract}
Malware is an emerging and popular threat flourishing in the underground economy. 

The commoditization of Malware-as-a-Service (MaaS) allows criminals to obtain financial benefits at a low risk and with little technical background. One such popular product is ransomware, which is a popular malware traded in the underground economy.

In ransomware attacks, data from infected systems is held hostage (encrypted) until a fee is paid to the criminals. This \textit{modus operandi} disrupts legitimate businesses, which may become unavailable until the data is restored, thus causing additional financial and reputational losses. A recent blackmailing strategy adopted by criminals is to leak data online from the infected systems if the ransom is not paid before a given time, threatening businesses to have their data exposed online. Besides reputational damage, data leakage might produce further economical losses due to fines imposed by data protection laws, e.g. GDPR in Europe. Thus, research on prevention and recovery measures to mitigate the impact of such attacks is needed to adapt existing countermeasures to new strains.

In this work, we perform an in-depth analysis of Avaddon, a ransomware offered in the underground economy as an affiliate program business. 

This threat has been linked to various cyberattacks and has infected and leaked data from at least 23 organizations. Additionally, it also runs Distributed Denial-of-Service (DDoS) attacks against victims that do not pay the ransom.
We first provide an analysis of the criminal business model from the underground economy. Then, we identify and describe its technical capabilities, and dissect details of their inner structure.
We provide empirical evidence of links between this variant and a previous family, suggesting that the same group was behind the development and, possibly, the operation of both campaigns. 
As a result, we provide tools to assist analysis, allowing to decrypt and label encrypted strings observed in the ransomware binary. 
Finally, we describe a method to decrypt files encrypted with Avaddon in real time. 
We implement and test the decryptor in a tool that can recover the encrypted data from an infected system, thus mitigating the damage caused by the ransomware. The tool is released open-source so it can be incorporated in existing Antivirus engines.
\end{abstract}


\end{frontmatter}


\section{Introduction} \label{ch:introduccion}

In February, 2018, the USA government estimated that cybercrime costs raised up to between 57 and 109 billions of dollars in 2016~\cite{CEA2020}. 
Cybercrime has been growing for the last decades as it becomes more profitable. The most common goal for cybercriminals is monetary gain and they commonly organize  to form online criminal enterprises and businesses~\cite{Collier20}. 
The virtual battlefield where such criminal activities operate allows miscreants to perpetrate crimes in countries with different extradition laws than those of the country where they reside. 
This strategy frequently makes cybercrime hard to prosecute, in addition to other technical characteristics that difficult attribution \cite{ODNI2018, InfosecInstitute}. 
In recent times, the underground economy has developed a myriad of approaches that allow cybercriminals to acquire high financial profits. 
With the cybercrime growth and specialization, many cybercriminals offer their products in an ``as-a-service'' model, where the attacker can purchase the service through the internet with little technical knowledge. 
These services reduce the entry level for new criminals and motivate newcomers into the underground ~\cite{Huang2018,Pastrana18}.

In 2017, Panda security analyzed around 15m binaries~\cite{Panda2018}. The most noticeable thing was that, upon reviewing the data collected, they realized that 99.1\% of the samples were only seen once, probably due to binary packing and encryption. Indeed, a common digital commodity offered in underground markets is malware~\cite{van2018}. Concretely,
one of the most popular variants offered is ransomware~\cite{Pwc2020}, where the attacker denies access to the data of its victims until a ransom is paid (hence the name of the threat). When these attacks affect companies or public organizations, they might provoke business interruptions, thus increasing the economic and social damage~\cite{Ghafur19}. Ransomware operators often partner up with other criminal groups, either in a customer-service relationship (offering the software for a fixed fee or via a subscription-based access to constant updates) or in a profit sharing scheme (one party is responsible for developing and maintaining the ransomware while the other distributes it, both sharing an arranged percentage of the revenues). Previous works show that criminals can run ransomware campaigns with little technical knowledge, making use of the available services, with an estimated return of investment of between 504\% and 12,682\%~ \cite{Huang2018}.

Due to the profitability and specialization of cybercrime, modern ransomware campaigns have improved their sophistication. First, techniques from well-established cryptography schemes, so-called hybrid cryptosystems, have been recently adopted in ransomware operations, combining symmetric and asymmetric cryptography. Second, modern ransomware perpetrators have incorporated another monetizing technique that further pushes the victims to pay a ransom: data leakage extortion. Apart from encrypting the files, ransomware operators now steal data from the infected systems and threaten the victims to leak it online if no ransom is paid. This extortion scheme was initiated by a threat actor named \textit{TWISTED SPIDER} in the last quarter of 2019 \cite{CrowdStrike2020}, and was quickly followed by other ransomware groups~\cite{Panda2020, Catalin2020, BankInfoSecurity20202}. In order to face the ransomware threat, and to be able to recover the hijacked files, it is important to understand the criminal ecosystem and also how the malware evolves and operates.

In this work, we study a novel ransomware campaign, dubbed Avaddon, which was launched on June 2020 in underground forums as a Ransomware-as-a-service (RaaS). Since then, Avaddon has been linked to various cyberattacks in 2020, and incorporates a recent trend on Ransomware which is is to publicly \textit{`blame and shame'} victims that do not pay the ransom~\cite{Intel471}. At the time of this writing, more than 574GB of data from 23 companies have been leaked and exposed online\footnote{For ethical and legal reasons, we have not downloaded nor checked the veracity of the exposed data since otherwise this would cause additional harm to users, and such analysis is not of public interest for the community~\cite{Thomas17}}. In addition, Avaddon operators have recently started to blackmail victims by running DDoS attacks. Existing reports described different technical features of Avaddon \cite{SubexSecure2020, idransomware2020, HornetSecurity2020, TrendMicro2020_1}. However, as far as the authors know, no public decryption procedure is available to recover files from an infection. We aim at filling this gap by providing an in-deph analysis of Avaddon and proposing a decryption routine that can decrypt files in real time, thus minimizing the impact of such attack. In particular, we analyze one of the first variants observed in early June, although the proposed decryption method is still functional for the latest samples of Avaddon at the time of this writing. The main contributions of this work are the following:

\begin{itemize}
    \item We analyze the Avaddon business ecosystem and provide an step-by-step analysis of its technical capabilities, using advanced static and dynamic analysis. This analysis can be generalized to grasp an overview of how modern ransomware operates, since their \textit{modus operandi} is similar. As a result of our study, we provide a set of indicators of compromise (IoCs) that may serve security analysts to develop further tools and countermeasures to detect Avaddon, such as signatures or heuristics.
    \item We showcase a typographical error in the list of services that Avaddon checks to avoid re-infecting victims, which is also present in modern variants of another Ransomware family, i.e., MedusaLocker. Additionally, we highlight that some similarities on the code of both families hint that they are operated or developed by the same group.
    \item We describe a method to recover the symmetric keys used for the encryption, thus allowing victims to recover the files from infected systems. Accordingly, we present and make publicly available a tool which can help victims to recover from these attacks in real time. While this tool was designed using the analysis of the first versions of Avaddon, we have confirmed that it still works with the most up-to-date versions of the ransomware, released in mid-January 2021. 
\end{itemize}

The rest of this work is structured as follows. Section~\ref{ch:background} describes the Avaddon criminal ecosystem and how it operates and evolves in the underground economy. Next, Section \ref{sec:analysis} discusses the results obtained after reverse engineering the sample using static and dynamic analysis and provides details of the internals of the ransomware. Section \ref{ch:key_recovery} describes the method to recover the session key used to encrypt the system and describes the remediation tool to decrypt the infected files. We provide experimental results in Section~\ref{sec:results} by infecting a sandboxed environment and decrypting the file system with the proposed approach. Finally, we conclude and discuss the implications and limitations of this paper in Section~\ref{ch:conclusions}.

\section{Background and related work} \label{ch:background}

In this Section, we first provide background information about how Ransomware works and existing defensive mechanisms. Then, we present the criminal ecosystem behind Avaddon, including its evolution in the underground economy, and how this has been reflected in the wild, i.e., leading to real-world cyberattacks.

\subsection{The Ransomware threat}
Ransomware is a type of malware that interrupts the business of the victim or denies access to its data until a ransom is paid, by means of data encryption. This type of malware has direct financial implications and has promoted the growth of cybercrime, where it is employed as a profitable business model \cite{Brewer2016}.

Before the popularization of cryptocurrencies, such as Bitcoin, online payment methods were risky for malware authors. SMS text messages, pre-paid cards or premium rate telephone numbers could be traced back easier than Bitcoin \cite{Zetter2015}. With the use of Bitcoin or other cryptocurrencies to ask for ransoms, it became much harder to trace the payments sent to criminals. Still, the characteristics of some cryptocurrencies allow for tracking transactions (although not connecting them to the attacker). For instance, Huang \textit{et al.} were able to track over \$16 million in likely ransom payments made by 19,750 potential victims during a two-year period \cite{Huang20182}. Thus, criminals have adopted privacy-preserving cryptocurrencies such as Monero that hinder tracking~\cite{Pastrana2019}. These cryptocurrencies, in combination with the cybercrime specialization, have promoted the ransomware threats as a profitable business for cybercriminals~\cite{Richardson2017}.

Ransomware detection approaches often leverage classical malware detection methods adopted for ransomware-specific behaviors. In this way, ransomware activities can be split in 8 stages~\cite{Hull2019}: \textit{fingerprint}, \textit{propagate}, \textit{communicate}, \textit{map}, \textit{encrypt}, \textit{lock}, \textit{delete} and \textit{threaten}. Kharaz \textit{et al.} focused the detection on common tasks performed by ransomware, such as changing the desktop wallpaper~\cite{Kharaz2016}. Some efforts have also been made to capture cryptography keys at runtime in order to facilitate decryption of infected systems~\cite{Kolodenker2017}. Following recent trends in malware detection, some Machine Learning-based approaches have also been proposed specifically targeting ransomware detection~\cite{Sgandurra2016, Vinayakumar2017, Lee2019}. 

\subsection{The ecosystem of Avaddon}
Avaddon\footnote{The name of the ransomware, Avaddon, may be derived from the Hebrew term Abaddon, the name of an angel of the abyss in the Bible, mainly associated with the meaning of ``destruction''~\cite{idransomware2020}.} is a ransomware that was offered as an affiliate program on June, 2020 in a Russian underground forum, only accessible by invitation or after the payment of a registration fee. Concretely, the operators were looking for partners for their campaign. Additionally, Avaddon was promoted on other underground forums afterwards.\footnote{Due to ethical reasons, and to avoid promoting the site, we do not provide the name of the forums.} Actors that become affiliates are equipped with both the ransomware binary and an administration panel from where they can control their infections. Access to the program is free and constrained only for reputed (and Russian-speaking) actors. In exchange for this, partners have to share part of the obtained revenues from the ransomware to the owners and operators. This share depends on the amount of infections, starting from 35\% and decreasing up to 15\% for larger volumes. Therefore, affiliates, who are only responsible for distributing and installing the malware on infected systems, gain 65\% of the revenues generated by the ransomware, without the need of operating the payment system~\cite{SubexSecure2020}. Such distribution often relies on botnets hired in a Pay-Per-Install scheme~\cite{Caballero2011}. Additionally, partners can purchase installs on RDP servers, which is another popular product traded in underground economy~\cite{KasperskyRDP16}. Thus, the supply chain needed to enter in this business does not require technical knowledge and it opens the barrier to any criminal entrepreneur~\cite{Bhalerao19}. 
As a restriction in the affiliates program of Avaddon, it is forbidden to target victims in the Commonwealth of Independent States (CIS). We describe the mechanism used to achieve this restriction in Section \ref{subsec:language_checks}. 

A few days after their publication on underground forums (on 2020-06-04 at around 14:00:00 UTC) Avaddon was observed in the wild. Concretely, it was distributed via mail in a malspam campaign~\cite{HornetSecurity2020}, 
that consisted in low-quality phishing emails that attached a malicious file. These emails hinted that a compromising photo of the victim had been leaked, inciting the victim to open the file out of fear. The attached file was a zip-compressed JavaScript file. This file tried to masquerade as a JPG photo, having the extension ``.jpg'' just before the ``.js'' extension (e.g., ``IMG123456.jpg.js''). Upon execution, the malicious file would download and execute the ransomware. 
Allegedly, the first wave of this campaign targeted mostly Canada, concretely various education services~\cite{HornetSecurity2020}. 
However, their targets varied later.
Indeed, as mentioned before, Avaddon was launched as a RaaS, which means that the targets are not chosen by the ransomware developers (apart from the ban on CIS victims) but by the affiliates.
Upon infecting a system, a ransom note is left to the victim with instructions on how to pay the ransom. The note would lead the victim to a Tor hidden service, where payment must be done in exchange for the decryptor. At the time of this writing, the payment service is still operative, confirming that the campaign is ongoing.

Shortly after Avaddon was first seen in the wild, Trend Micro conducted and released a technical analysis~\cite{TrendMicro2020_1}. The report offers an overview of the ransomware capabilities and \textit{modus operandi}. However, no decryption option is mentioned (it only indicates how to remove the ransomware). Regarding the decryption process upon paying the ransom, some stories from affected users state that it is unreliable and recovery is not ensured~\cite{DecryptStory12020}. 

Two months after the initial release, on August, 2020, Avaddon was updated to incorporate a new trending technique to their features~\cite{TrendMicro2020_2}: extortion to victims. Following the model from other ransomware campaigns, Avaddon operators decided to threat victims to exfiltrate their data, by making it publicly available if they do not pay the ransom~\cite{BankInfoSecurity2020, BleepingComputer2020}. By the end of January, 2021, Avaddon has allegedly infected and leaked full dump data from 20 companies (totalling 574.46 GB of data) and is extorting (i.e., threatening of leaking data) 3 other companies which have been recently infected.
Finally, in January 2021 (concurrent to the writing of this paper), Avaddon included a new technique used for extortion: attacking their victims with Denial-of-Service Attacks~\cite{BleepingComputerDDoS}. Therefore, the threat to victims is now three-fold, i) data is first encrypted in the infected systems, so it becomes unavailable, ii) data is leaked publicly if the ransom is not paid, and iii) a DDoS attacks is performed to disrupt their businesses until the ransom is paid. 

At the time of writing, we are not aware of any public decrypting tool for Avaddon. Additionally, various reports and recent complaints from Avaddon victims about their decryption support~\cite{AmigoA2020, PintSizeNore2020} show that the campaign is still operative.
In this paper, we fill this gap and release an open-source tool that automatically detects and decrypts files, which could be integrated in existing Antivirus solutions. 

\section{Ransomware analysis}\label{sec:analysis}

In this section, we provide an analysis of the Avaddon ransomware, concretely one released as part of their initial advertising campaign on June 2020. We follow standard techniques for malware analysis, concretely static and dynamic analysis. To perform the aforementioned analysis, we utilize popular tools for binary analysis (i.e., Binary Ninja~\footnote{https://binary.ninja/}, x64dbg~\footnote{https://x64dbg.com/} and Pestudio~\footnote{https://www.winitor.com/}) and a virtual machine to run the ransomware safely. We build the virtual environment on top of VirtualBox~\footnote{https://www.virtualbox.org/} and install Windows 7 x64 in the virtual system. 

The analyzed binary (MD5:\hash{c9ec0d9ff44f445ce5614cc87398b38d}) is a Portable Executable (PE) file.
The PE format describes the structure of executable programs in Windows Operating Systems (OS)~\cite{MSPEFormat}. PE files are mainly divided in two important pieces: headers and sections. While headers contain information about the program itself and data to be read by the OS in order to correctly load and execute the program, sections contain the actual code and data of the program. Additionally, we see that its size is 1.1 MB, so it is not a large program. Finally, we see that the compilation time field of the binary is set to June, 3, 2020, at 11:47:22 (UTC). Although this field is prone to be modified by malware authors in order to confuse analysts, the timestamp is similar to the time of the first appearances of Avaddon samples \cite{SubexSecure2020}, which confirms that we are indeed analyzing one of the first versions of Avaddon.

We describe the packing protections of the analyzed binary in Section \ref{subsec:packing_protections}. Next, we show the imported functions and the extracted strings in Sections \ref{subsec:imports} and \ref{subsec:strings}, respectively. In Section \ref{subsec:anti_analysis}, we show the anti-analysis techniques employed by the binary. Then, we show how the ransomware authors implemented a protection to not infect Commonwealth of Independent States (CIS) victims in Section \ref{subsec:language_checks} and analyze the privilege escalation techniques, step by step, in Section \ref{subsec:privilege_escalation}. The details of the persistence mechanism are showcased in Section \ref{subsec:persistence}. Following, interactions with other processes and services are presented in Section \ref{subsec:services_manipulation}. Finally, we expose the cryptography mechanisms used in the last two sections, concretely key management (Section \ref{subsec:key_generation}) and file encryption (Section \ref{subsec:file_encryption}).

\subsection{Packing protections} \label{subsec:packing_protections}

Looking at some properties of the PE file, we conclude that the sample is not packed. First, we find that the PE file contains 4 sections which have almost no differences in size between disk and memory. This in an indicator of the PE file not being packed, since the presence of a virtual section (i.e., a section that requests space in memory but does not occupy bytes in disk) is a common indicator of packing protections. Then, we find over 200 imported functions and several meaningful strings, which present some useful information about the capabilities of the ransomware. Often, packing protections attempt to hide imports and strings in order to difficult analysis. Finally, the entropy levels of the PE file are not high enough to hint the presence of a packer. The highest entropy level is reached in the ``.text'' section, whose entropy value is 6.62. However, this does not confirm the existence of packed or encrypted code, which often have values ranging from a minimum of 6.677 to a maximum of 6.926 \cite{Lyda2007}.

\subsection{Imported functions} \label{subsec:imports}

The Windows OS offers an Application Programming Interface (API) which abstracts many functionalities from developers, e.g to interact with files, processes, etc. This also provides an abstraction layer regarding the underlying hardware. In order to call those functions, programs need to know their location in memory. This need may be fulfilled in different ways, but the most common method consists on importing the required functions prior to execution. This is done by the OS loader before transferring control to the program. To do so, the PE file contains an Import Address Table (IAT) in the headers, which includes a list of functions to be imported by the OS loader. When the file is executed, the OS loads the file in memory and fills the IAT with the addresses of each requested function. Then, the program is able to call those functions because it now knows where each of these functions is allocated in memory. Therefore, the IAT provides useful information about the capabilities and intentions of the program. 
The functions imported by the Avaddon sample analyzed show common capabilities of ransomware, such as encryption (e.g., \textit{CryptGenKey} or \textit{CryptEncrypt}), persistence (e.g., \textit{RegCreateKeyW}, \textit{StartServiceW}), anti-analysis (e.g., \textit{IsDebuggerPresent}) or activity control (e.g., \textit{DeleteService} or \textit{TerminateProcess}).

\subsection{Strings} \label{subsec:strings}

Looking for strings through a PE file allows analysts to identify capabilities of the binary, as well as looking at the IAT. Indeed, some imports will appear when searching for strings if they are imported by name (external functions may be imported by name or ordinal \cite{MSPEFormat}). Therefore, we proceed to extract all readable strings that have more than 4 characters in the whole file. Then, we filter the extracted strings and exclude those that are not meaningful (bytes that are part of code may non-intentionally form readable strings that are not meaningful). In this case, as aforementioned, we find enough meaningful strings to think that the PE file is not packed. Many of the strings found were paths to folders or files (e.g., ``\hash{C:\textbackslash Temp}''). While we initially can not know the actual purpose of those files, we hypothesize that some of them may be used to drop additional payloads or to move the PE file upon infection to a different location (we confirm this hypothesis in Section \ref{subsec:persistence}).

Among the strings that are present in the PE file, we observe two of them that refer to cryptography providers (i.e., ``\hash{Microsoft} \hash{Enhanced} \hash{Cryptographic} \hash{Provider} \hash{v1.0}'' and ``\hash{Microsoft} \hash{Enhanced} \hash{RSA} \hash{and} \hash{AES} \hash{Cryptographic} \hash{Provider}''). These strings are normally used to acquire cryptography contexts using the Windows API, which are later needed to perform some cryptography operations. Additionally, some strings indicate that the ransomware was developed in C++. C++ is an object-oriented language.  Although this characteristic does not provide any information about the capabilities of the sample, the particularities of C++ programs must  be  taken  into  account  in  the  analysis  process. We will  highlight some C++ properties that allow us to extract conclusions in the reverse engineering process, but discussing the differences between C++ and other languages at assembly level is out of the scope of this work. For more information, we refer to Sabanal \textit{et al.} \cite{sabanal2007}.

Interestingly, we find many strings that are Base64 encoded. However, upon decoding them, no legible string is recovered. Therefore, we suspect that these strings are obfuscated by other means (i.e., encoding or encryption) in order to hide their content. This is a common mechanism in malware samples. In such case, those strings may be of importance to understand additional capabilities of the malware that may not be retrieved without further analysis. 
We thus confirm that these strings are indeed encrypted and are only decrypted at runtime on demand, i.e., when they are required by the program. First, global variables are created to hold the encrypted strings, making them accessible from every function in the binary. In Algorithm \ref{alg:initialize_global_var}, we show one of the functions (0x4012a0 in this case) that creates a global variable pointing to an encrypted string. There, the encrypted string and its size are pushed onto the stack (lines 1-2). Then, a global variable is created at 0x4f8a28 with the content of the encrypted string (lines 3-4). Finally, a destruction function is registered (lines 5-6). This function will be called when the process exits. The global variable is then referenced wherever this particular string is needed in the program, and each encrypted string has its own initialization function. These functions are the constructors of the global variables. We know that these are global variables in the source code because:

\begin{enumerate}
    \item There is one global variable per encrypted string and one constructor function for each global variable.
    \item Each global variable has a predefined address. These addresses are hardcoded in each constructor function.
    \item After initializing the global variable, a destructor function is registered to be called upon terminating the program.
\end{enumerate}

\begin{algorithm}
        \SetAlgoLined
        \caption{One of the functions responsible for initializing a global variable with the value of an encrypted string.} 
        \label{alg:initialize_global_var}   
        0x4012a0: push 0x30\; \tcp{Size of the encrypted string}
        0x4012a2: push 0x49e180\; \tcp{Encrypted string}
        0x4012a7: mov ecx, 0x4f8a28\; \tcp{Global variable}
        0x4012ac: call 0x40a390\; \tcp{Creates a global variable at ecx (0x4f8a28 in this case) with the string stored at the value previously pushed (0x49e180 in this case)}
        0x4012b1: push 0x4874a0\; \tcp{Destructor}
        0x4012b6: call \_atexit\; \tcp{Register the destructor function to be called when the process ends}
        0x4012bb: pop ecx\;
        0x4012bc: retn\;
\end{algorithm}

Once initialized, the strings are decrypted and used as needed by referencing the global variables. In Algorithm \ref{alg:use_decrypted_string}, we show an example of this procedure. In particular, a string containing some command line arguments is decrypted and used to create a process. In this case, the goal is to delete security backups. First, the decryption function is called passing the global variable as an argument (lines 1-3). This function returns a new string with the decrypted value, which is immediately used to create the aforementioned process (lines 4-5). The sequence of instructions described can be summarized in the following pseudo code:

\begin{itemize}
\item[]\textit{CommandLine = DecryptString(GlobalVariable);}
\item[]\textit{CreateProcess(CommandLine);}
\end{itemize}

\begin{algorithm}
        \SetAlgoLined
        \caption{Decryption of a global variable into a temporary register.} 
        \label{alg:use_decrypted_string}   
        0x40d110: mov     edx, 0x4f8a28\; \tcp{Global variable that contains an encrypted string}
        0x40d115: lea     ecx, [esp+0x8]\; \tcp{Local variable that will hold the decrypted string}
        0x40d119: call    decrypt\_string\; \tcp{Decrypts the string at edx (the global variable) and stores the result in ecx (the local variable)}
        0x40d11e: push    eax\; \tcp{eax now contains the decrypted string (it is equal to [esp+0x8], the local variable) which, in this case, is a command line}
        0x40d11f: call    create\_process\; \tcp{Creates a process with the command line received as argument}
\end{algorithm}

The decryption function is located at address 0x40c780. First, the received string is decoded from Base64. Then, as shown in Algorithm \ref{alg:decrypt_string}, each character is decrypted by substracting 2 units from its value (line 3) and XOR-ing the result with 67 (line 5). These instructions are executed once for every character in the string.

\begin{algorithm}
        \SetAlgoLined
        \caption{Characters decryption.} 
        \label{alg:decrypt_string}   
        0x40c820: mov     al, byte [esi]\; \tcp{Move the current character to al (the lower 8 bits of eax)}
        0x40c822: mov     edx, dword [ebp-0x1c]\;
        0x40c825: sub     al, 0x2\; \tcp{Substract two units from the character}
        0x40c827: mov     edi, dword [ebp-0x18]\;
        0x40c82a: xor     al, 0x43\; \tcp{XOR the result with 0x43}
        0x40c82c: mov     byte [ebp-0x30], al\;
        0x40c82f: cmp     edx, edi\;
        0x40c831: jae     0x40c84d\;
\end{algorithm}

Since we know the address in which global variables are placed, we have automatically re-labeled them in order to improve the readability of the code for the analyst. To allow for reproducibility and to assist other analyses on this and similar malware samples, we publish a script that automates these tasks using the \textit{Binary Ninja} tool in our public repository~\footnote{\url{https://github.com/JavierYuste/AvaddonDecryptor}}.

\subsection{Anti-analysis techniques} \label{subsec:anti_analysis}

Successfully infecting a system critically depends on not being detected. Thus malware authors often implement different techniques to evade antivirus systems or sandboxes. Additionally, mechanisms are frequently put in place in order to delay analysts and, therefore, increment the time needed for building detection tools for the sample (e.g., signatures). In the case of Avaddon, the binary is not packed, which is a common obfuscation technique. However, we observe other anti-analysis techniques, described next. 

\textbf{String obfuscation}. As mentioned in prior sections, various of the strings are encrypted, which may hide important functionality. This technique is commonly used to: i) evade detection, and ii) delay analysts. See Section~\ref{subsec:strings} for a detailed description of this obfuscation technique and the process used to decrypt the strings.

\textbf{Anti-debugging}. We found a call to \textit{IsDebuggerPresent} at offset 0x42e03d. Debuggers are programs designed to analyze other programs at runtime (i.e., processes), and are used by security analysts to dynamically inspect malware. Hence, malware authors often embed code in their programs that checks for debuggers and, if detected, terminates the process or changes their behavior. In particular, \textit{IsDebuggerPresent} is a function provided by the Windows API. 
If a debugger was attached to the program, this function would return true and the binary would exit. To circumvent this protection, we consider two options:

\begin{enumerate}
    \item Hook the call to \textit{IsDebuggerPresent} so it always returns false. By doing this, we bypass any check done by the malware, changing the code on the fly, and the debugger would not be detected by the sample.
    \item Change a binary value in the the Process Environment Block (PEB), a data structure that holds information about the process. That structure is built by the OS when executing the program and is unique per process. Among other information, it contains a bit that indicates if a debugger has been attached. When a call to \textit{IsDebuggerPresent} is made, it returns the value of that bit. Therefore, changing the value in the PEB would successfully hide the debugger from that call and from any manual checks (the PEB may also be walked through manually by parsing its structure). 
\end{enumerate}

In order to avoid further anti-debugging mechanisms that may parse the PEB (i.e., not using \textit{IsDebuggerPresent}), we decided to implement the second option.

\subsection{Language checks} \label{subsec:language_checks}

To avoid infecting systems in some countries, it is frequently observed that malware binaries implement techniques to check the country where the infected machine is located, so as to ensure that citizens from some regions are not affected. It is common to see that CIS victims are dodged in many malware samples, as it is the case for this one. The most popular approach is to check for the keyboard layouts and the OS language. In this sample, we found both checks for different layouts and languages (addresses 0x42e0ec and 0x42e0b6, respectively). In particular, we discovered checks for language locales (i.e., Russian and Ukrainian) and keyboard layouts (i.e., Russian, Sakha, Tatar and Ukrainian). If any of these keyboard layouts or OS locales is found, the binary exits without harming the landed system. That is, this sample of Avaddon ransomware is designed to avoid infecting Russian and Ukrainian systems. This, together with the fact that the malware was first advertised in a Russian underground forum, provides strong (though not conclusive) evidence that the origin of the malware is Russia.

\subsection{Privilege escalation} \label{subsec:privilege_escalation}
Malware authors often spend great resources in order to infect systems, e.g. to gain initial access and evade detection by AV software. However, having invested so much effort in those tasks, their immediate post-infection activities might fail due to the need for administrator privileges if the user becomes suspicious after being requested to concede those privileges. Therefore, reducing the number of clicks needed from the victim is critical.
Indeed, malware actions usually require administrator privileges in the infected system to accomplish some critical tasks (e.g., acquire persistence, infect system files or processes, etc.). In this particular case, escalating privileges is critical because the ransomware needs to i) acquire persistence through registry keys (Section \ref{subsec:persistence}), ii) stop processes and services (Section \ref{subsec:services_manipulation}), and iii) delete backups (Section \ref{subsec:file_encryption}).

The process implemented to elevate privileges in Avaddon is a well known User Account Control (UAC) bypass. Indeed, there are public open-source implementations~\cite{hfiref0x2017} and it is not uncommon to find this technique in different malware families~\cite{Morphisec2020, SecurityInBits2019}. Next, we briefly summarize this process and how it is implemented in Avaddon. First, three registry keys are added or modified (at offset 0x40ed20). Concretely, these keys are:

\begin{enumerate}
    \item \hash{HKEY\_LOCAL\_MACHINE\textbackslash SOFTWARE\textbackslash Microsoft\textbackslash Windows\textbackslash CurrentVersion\textbackslash Policies\textbackslash System} \hash{EnableLUA = 0} (disables the ``administrator in Admin Approval Mode'' user type \cite{MSEnableLua})
    \item \hash{HKEY\_LOCAL\_MACHINE\textbackslash SOFTWARE\textbackslash Microsoft\textbackslash Windows\textbackslash CurrentVersion\textbackslash Policies\textbackslash System} \hash{ConsentPromptBehaviorAdmin = 0} (this option allows the Consent Admin to perform an operation that requires elevation without consent or credentials \cite{MSConsentPrompt})
    \item \hash{HKEY\_LOCAL\_MACHINE\textbackslash SOFTWARE\textbackslash Microsoft\textbackslash Windows\textbackslash CurrentVersion\textbackslash Policies\textbackslash System} \hash{EnableLinkedConnections = 1} (makes user mapped drives available to the administrator versions of those users \cite{MSLinkedConn})
\end{enumerate}

The first two registry key values allow the sample to elevate privileges without alerting the user, and the third enables the access to volumes of the current user when administrator privileges are acquired.

Then, the sample checks its privileges (offset 0x41a5c0). If it has administrator privileges, it continues its execution without running the rest of the UAC bypass. Otherwise, administrator privileges are obtained by running the following procedure (implemented at 0x40ef90):

\begin{enumerate}
    \item First, a class ID (CLSID) is decrypted. This CLSID was stored in the binary as an encrypted string, as we described in Section \ref{subsec:strings}. The decrypted value is ``\hash{\{3E5FC7F9-9A51-4367-9063-A120244FBEC7\}}'', which corresponds to CMSTPLUA. For the rest of this section, we refer to it as CLSID\_CMSTPLUA.
    \item Next, an IID is decrypted in the same way, obtaining the value ``\hash{\{6EDD6D74-C007-4E75-B76A-E5740995E24C\}}''. For the rest of this section, we refer to it as IID\_ICMLuaUtil.
    \item Then, a third string is decrypted, which contains the value ``\hash{Elevation:Administrator!new:}''.
    \item Once the three strings have been decrypted, a new string is built by concatenating ``\hash{Elevation:Administrator!new:}'' and CLSID\_CMSTPLUA.
    
    \item Next, the execution calls the function \textit{CoGetObject} in order to obtain a pointer to CMLuaUtil. The parameters of the call are as follows:\\
    
    \quad \hash{CoGetObject(}``\hash{Elevation:Administrator!new:\{3E5FC7F9-9A51-4367-9063-A120244FBEC7\}}''\hash{,} \hash{0x24,} \hash{\&IID\_ICMLuaUtil,} \hash{\&CMLuaUtil)}\\
    
    At this point, user interaction might be needed to grant administrator privileges for the program in some systems. In some cases, this might be accompanied by social engineering techniques, e.g. instructions accompanying the phishing email where the malware is attached. In this case, we have not observed any particular behavior.
    
    \item If the call is successful, CMLuaUtil now points to a structure that contains the address of a function named \textit{ShellExec} (CMLuaUtil$\xrightarrow{}$lpVtbl$\xrightarrow{}$ShellExec). 
    
    \item After obtaining the absolute path of the malware PE file (via a call to \textit{GetModuleFileNameW}) the binary executes itself with administrator privileges by calling \textit{ShellExec} with the following parameters:\\
    
    \quad \hash{ShellExec(CMLuaUtil,} ``\hash{C:\textbackslash [...]\textbackslash sample.exe}''\hash{,} \hash{[...])}
\end{enumerate}

\subsection{Persistence and infection tracking} \label{subsec:persistence}

In order to survive across reboots, malware samples must be run automatically on infected systems after the initial foothold has been obtained~\cite{Lockheed2020}. Otherwise, they would need to infect the system again if further runs are required. In order to achieve persistence in the system, there exist many approaches. Usually, malware authors acquire persistence by adding registry keys, creating services or registering scheduled tasks. By doing so, the malware sample is automatically run by the OS (e.g., at scheduled times or at every reboot).
Additionally, malware samples often implement mechanisms to prevent re-infection of already-infected systems, thus to minimize the risks of detection or to prevent disruption of previous runs.

By looking at the imported functions (see Section \ref{subsec:imports}) we hypothesize that persistence may be acquired via registry keys or services. Then, using dynamic analysis we confirm that persistence is obtained by adding two registry keys. Upon inspecting the code of the binary, we locate the function responsible for acquiring persistence at address 0x40cf50. The only purpose of this function is to add the following registry keys:

\begin{itemize}
    \item \hash{HKU\textbackslash S-1-5-21-2724635997-1903860598-4104301868-1000\textbackslash Software\textbackslash Microsoft\textbackslash Windows\textbackslash CurrentVersion\textbackslash Run\textbackslash update:} \hash{"C:\textbackslash Users\textbackslash \%User Profile\%\textbackslash AppData\textbackslash Roaming\textbackslash \%sample\%.exe"}
    \item \hash{HKLM\textbackslash SOFTWARE\textbackslash Wow6432Node\textbackslash Microsoft\textbackslash Windows\textbackslash CurrentVersion\textbackslash Run\textbackslash update:} \hash{"C:\textbackslash Users\textbackslash \%User Profile\%\textbackslash AppData\textbackslash Roaming\textbackslash \%sample\%.exe"}
\end{itemize}

With those registry keys in the system, the PE file is executed at each system reboot (notice that a copy of the sample is dropped at runtime in \hash{"C:\textbackslash Users\textbackslash \%User Profile\%\textbackslash AppData\textbackslash Roaming\textbackslash \%sample\%.exe"}, where ``\hash{\%sample\%}'' is the name of the PE file). 
To avoid re-infecting a system more than once, a mutex is created with the value \hash{\{2A0E9C7B-6BE8-4306-9F73-1057003F605B\}}. If this mutex is already present in the system, the binary exits and does not encrypt files. In addition, the ransomware takes measures to avoid encrypting already encrypted files, as we describe in Section \ref{subsec:file_encryption}. Thus, having mechanisms to prevent re-infection of a machine might be to avoid reinfecting victims that have already payed a ransom. Nevertheless, the fact that the presence of such mutex is checked allows to prevent Avaddon infections. By creating such mutex in a healthy system, Avaddon ransomware samples will not execute, acting as an Avaddon vaccine. However, not every sample of Avaddon uses the same mutex, as it may change among versions.

\subsection{Process and service manipulation} \label{subsec:services_manipulation}

In order to avoid being detected or neutralized, some malware samples try to stop anti-malware solutions. In order to do so, administrator privileges must be acquired. However, it is often easier to acquire administrator privileges without being detected than to encrypt the whole file system without rising awareness. In Section \ref{subsec:imports}, we highlighted that the PE file imported some functions that may indicate an attempt to control some anti-malware solutions by interacting with services and processes. Additionally, before attempting to encrypt files, it is important to stop processes that may be locking some files. For instance, ransomware authors may look to stop database processes that may be locking database files.

In this case, we find two functions (located at offsets 0x41a8f0 and 0x40c990 of the sample) that try to stop a list of services and processes, respectively, if found in the system. As expected, among those lists, we have found anti-malware solutions (e.g., ``DefWatch'') and databases (e.g., ``sqlservr'').

We notice that the name of one of the services is misspelled. Concretely, ``vmware-usbarbitator64'' is missing an `r' (and should instead be ``vmware-usbarbitrator64''). This typographical error was found in other ransomware family, MedusaLockker. This indicates that developers reuse code from other families~\cite{TAU2020, Zsigovits2020}. We are unaware on whether this is due to the same actor developing both families, or due to code reuse from one to another (though we have found no evidence of the source code of MedusaLocker being leaked).
Indeed, we notice that the Tactics, Techniques and Procedures (TTPs) of Avaddon are very similar to those of MedusaLocker if we compare our analyses with the report on MedusaLocker from Carbon Black’s Threat Analysis Unit \cite{TAU2020}. This is an interesting fact regarding the attribution of this campaign which might require further investigation if future families share this peculiarity.

\subsection{Key generation} \label{subsec:key_generation}

One of the most critical parts of a ransomware campaign is the encryption process. The keys used, how they are imported or generated, how they are exported, the encryption algorithm chosen, etc., are important decisions for malware developers. An error in this process may allow analysts to develop measures to recover encrypted files, completely neutralizing the campaign revenues. In this case, two keys are used in the encryption process in a so-called hybrid scheme. One key (the session key) is randomly generated in each execution and used to encrypt the files in the system. This key is used in a symmetric encryption scheme, AES256. Therefore, the same key must be used to decrypt the affected files. The second key is a public one, part of an asymmetric scheme, RSA1. This key is imported (it is present in the PE file) and used only to encrypt the previously generated key. Therefore, the session key can only be decrypted by the malware authors, since the private key of the asymmetric scheme is only known by them.

The whole process that we described in the previous paragraph is split in three functions in the PE sample. These functions, responsible for key management, are located at offsets 0x413600, 0x413a60 and 0x413f50 respectively. 

\paragraph{Public key import} The function at 0x413600 is responsible for importing the public key. The import is made by calling the Windows API function \textit{CryptImportKey} with the following parameters:\\

\hash{CryptImportKey(hProv: CSP,} \hash{pbData:} \hash{Key} \hash{to} \hash{be} \hash{imported,} \hash{dwDataLen:} \hash{Length} \hash{of} \hash{the} \hash{key,} \hash{hPubKey:} \hash{0,} \hash{dwFlags:} \hash{0,} \hash{phKey:} \hash{Handle} \hash{to} \hash{the} \hash{imported} \hash{key} \hash{after} \hash{the} \hash{call)}\\

The key (which is Base64 encoded) is part of a RSA1 public/private pair. As per the documentation \cite{MSImportKey}, the parameter \textit{hPubKey} must be equal to 0 when the key to be imported is a public key (a \textit{PUBLICKEYBLOB} object). This detail indicates that the imported key is actually the public one of the pair.

\paragraph{Generated key} After importing the public key, a random key is generated. This randomly generated key (the session key) is used to encrypt the files of the system later, using an AES256 scheme. The function responsible of generating the session key is the one located at 0x413f50. To generate it, a function from the Windows API is called, \textit{CryptGenKey}, with the following parameters:\\

\hash{CryptGenKey(hProv:} \hash{CSP,} \hash{Algid:} \hash{CALG\_AES\_256,} \hash{dwFlags:} \hash{CRYPT\_EXPORTABLE,} \hash{phKey:} \hash{Handle} \hash{to} \hash{the} \hash{generated} \hash{key} \hash{after} \hash{the} \hash{call)}\\

The parameter \textit{Algid} indicates that the generated key is to be used in AES256. Additionally, notice that the flags passed to the function indicate that the key must be exportable. Once the key has been generated, it is exported and encrypted using the previously imported RSA1 key. The result is then included in the ransom note, in order to allow the ransomware operators to recover the encryption key and provide a decryption tool to those victims that decide to pay a ransom.

\paragraph{Keys destruction} Finally, the function located at 0x413f50 is the one responsible for securely destroying the keys. This function will destroy the public RSA1 key and the generated AES256 key. The purpose of this function is to ensure that they do not remain in memory after being used. However, this function is only called when the process exits, which only occurs when the infected system is shutdown (the ransomware process remains active to also encrypt new files). Therefore, the session key is never destroyed if the system is not powered off. This is a mistake from the malware perspective since, as long as the computer remains active, the key is kept in memory and thus can be retrieved using basic forensics techniques. In Section \ref{ch:key_recovery}, we will take advantage of this detail to describe and present a tool to recover the symmetric key generated and decrypt all the affected files.

\subsection{File encryption} \label{subsec:file_encryption}

In Section \ref{subsec:key_generation}, we presented the mechanism used to generate the key used to encrypt files. Additionally, we showed that the algorithm used to encrypt files is AES256, a symmetric encryption scheme. In this Section, we will describe the process followed to encrypt files in the infected system.

The first step performed by the ransomware is to delete backups so the original files cannot be restored by locally saved security copies. To achieve that goal, the function at 0x41a800 executes the following processes:

\begin{itemize}
\item \hash{wmic.exe} \hash{SHADOWCOPY} \hash{/nointeractive}
\item \hash{wbadmin} \hash{DELETE} \hash{SYSTEMSTATEBACKUP}
\item \hash{wbadmin} \hash{DELETE} \hash{SYSTEMSTATEBACKUP} \hash{-deleteOldest}
\item \hash{bcdedit.exe} \hash{/set} \hash{\{default\}} \hash{recoveryenabled} \hash{No}
\item \hash{bcdedit.exe} \hash{/set} \hash{\{default\}} \hash{bootstatuspolicy} \hash{ignoreallfailures}
\item \hash{vssadmin.exe} \hash{Delete} \hash{Shadows} \hash{/All} \hash{/Quiet}
\end{itemize}

In order to successfully execute those processes, administrator privileges are needed, which were obtained using the procedure that we described in Section \ref{subsec:privilege_escalation}. Finally, the contents of the recycle bin are deleted by calling the Windows API function \textit{SHEmptyRecycleBinW}.

Next, files are encrypted following a depth-first search approach. Microsoft SQL and Exchange folders are prioritized, being the first ones to be encrypted. Then, the root path is encrypted (i.e., \hash{C:\textbackslash \textbackslash *}). Finally, shared folders and mapped volumes are enumerated and encrypted (e.g., \hash{D:\textbackslash \textbackslash *}, \hash{Y:\textbackslash \textbackslash *}, or \hash{\textbackslash \textbackslash VBoxSvr\textbackslash \textbackslash shared\_folder\textbackslash \textbackslash *}). Therefore, the order in which folders are encrypted, following a depth-first approach, is the following:\\

\begin{enumerate}
    \item \hash{C:\textbackslash \textbackslash Program} \hash{Files\textbackslash \textbackslash Microsoft\textbackslash \textbackslash Exchange} \hash{Server\textbackslash \textbackslash *}
    \item \hash{C:\textbackslash \textbackslash Program} \hash{Files} \hash{(x86)\textbackslash \textbackslash Microsoft\textbackslash \textbackslash Exchange} \hash{Server\textbackslash \textbackslash *}
    \item \hash{C:\textbackslash \textbackslash Program Files\textbackslash \textbackslash Microsoft} \hash{SQL} \hash{Server\textbackslash \textbackslash *}
    \item \hash{C:\textbackslash \textbackslash Program} \hash{Files} \hash{(x86)\textbackslash \textbackslash Microsoft SQL Server\textbackslash \textbackslash *}
    \item \hash{C:\textbackslash \textbackslash *}
    \item Shared folders and mapped volumes
\end{enumerate}

For each file encountered, the process performs three checks before the actual encryption. 
\begin{enumerate}

\item \textbf{Strings from a whitelist}. The path is checked to not contain specific strings (see Appendix \ref{appedix:list_whitelisted_strings} for the list of skipped strings). If the absolute path of the file contains one of those strings, the file is left untouched. This check is excluded for the first four folders searched, those that belong to Microsoft SQL and Exchange servers. Therefore, this check is applied only to searches initiated at the root folder (i.e., C:\textbackslash \textbackslash *) or shared folders and mapped volumes. 
\item \textbf{File extensions}. The extension of the file is checked. The extensions that are excluded (not encrypted) are the following: \hash{bin}, \hash{ini}, \hash{sys}, \hash{dll}, \hash{lnk}, \hash{dat}, \hash{exe}, \hash{drv}, \hash{rdp}, \hash{prf}, \hash{swp}, \hash{mdf}, \hash{mds} and \hash{sql}.
\item \textbf{Prevent re-encryption}. The third test checks if the file has already been encrypted by Avaddon. To do so, a signature at the end of the file (that is left after encrypting a file by the ransomware, as we will describe later in this section) is read. In particular, the last 24 bytes of the file are read. If the file has been previously encrypted, it should contain the hexadecimal values 0x200 and 0x1030307 at offsets 8 and 16 in those 24 bytes.
\end{enumerate}

If none of these checks is positive then the file is encrypted. 
The encryption process is done by the function located at virtual address 0x413bb0. This function receives a copy of the AES256 key (see Section \ref{subsec:key_generation}) and the name of the file to be encrypted. We present a high-level pseudo code (some function signatures have been simplified to avoid using pointers) extracted from the analyzed function in Algorithm \ref{alg:file_encryption}. First, the size needed for the buffer to hold the bytes after encryption is calculated (line 1). Then, the file contents are read in chunks of 0x100000 bytes (line 5) and encrypted in blocks of 0x2000 bytes (lines 8-9). However, although there exists a loop to read and encrypt the whole file, only the first 0x100000 bytes are encrypted. This is due to the last call to \textit{SetFilePointerEx}, which sets the file pointer to the end of the file (line 18). When there are only 0x2000 or less bytes left to be encrypted (line 13), the last chunk of bytes is encrypted (lines 14-15) and written to the file (line 16). Notice that the parameter \textit{Final} (line 15) in the call to the encryption routine is always set to \textit{False}. This parameter should be \textit{True} if the block to encrypt is the last block of the file. We will need to take this detail into account in Section \ref{ch:key_recovery}. Finally, 512 unused bytes and the signature are written at the end of the file to mark it as encrypted (lines 20-22)\;

\begin{algorithm}
        \SetAlgoLined
        \caption{Function responsible for encrypting files.} 
        \label{alg:file_encryption}   
        \KwIn{\textit{File}, file to be encrypted\linebreak
                \textit{Key}, a duplicate of the AES256 key}
        \medskip
        \textit{buffer\_size} $\leftarrow$ CryptEncrypt(hKey: Key, Final: False, pbData: 0, pdwDataLen: 0x2000)\;
        \textit{file\_size} $\leftarrow$ GetFileSizeEx(hFile: File)\;
        \textit{file\_pointer} $\leftarrow$ 0\;
        \Do{\textit{number\_of\_bytes\_read} $\geq$ 0x100000 \&\& \textit{file\_pointer} $<$ \textit{file\_size}}{
            \textit{bytes\_read}, \textit{number\_of\_bytes\_read} $\leftarrow$ ReadFile(hFile: \textit{File}, offset: \textit{file\_pointer}, nNumberOfBytesToRead: 0x100000)\;
            \textit{i} $\leftarrow$ 0\;
            \Do{\textit{i} $\leq$ \textit{number\_of\_bytes\_read} - 0x2000}{
                \textit{bytes\_to\_encrypt} $\leftarrow$ \textit{bytes\_read[i:i+0x2000]}\; \tcp{The file is encrypted in blocks of 0x2000 bytes}
                \textit{encrypted\_bytes} $\leftarrow$ CryptEncrypt(hKey: \textit{Key}, Final: False, pbData: \textit{bytes\_to\_encrypt})\;
                WriteFile(hFile: \textit{File}, lpBuffer: \textit{encrypted\_bytes})\;
                \textit{i} = \textit{i} + 0x2000\;
            }
            \If{\textit{number\_of\_bytes\_read} - \textit{i} $<$ 0x2000}{
                \textit{bytes\_to\_encrypt} $\leftarrow$ \textit{bytes\_read[i:]}\;
                \textit{encrypted\_bytes} $\leftarrow$ CryptEncrypt(hKey: \textit{Key}, Final: False, pbData: \textit{bytes\_to\_encrypt})\;
                WriteFile(hFile: \textit{File}, lpBuffer: \textit{encrypted\_bytes})\;
            }
            \textit{file\_pointer} $\leftarrow$ SetFilePointerEx(hFile: \textit{File}, liDistanceToMove: \textit{0}, dwMoveMethod: \textit{FILE\_END}) \; \tcp{This call sets the file pointer to the end of the file. This is done to stop processing more bytes from the file}
        }
        WriteFile(hFile: \textit{File}, lpBuffer: \textit{VictimID})\; \tcp{The Victim ID is written to the end of the file}
        \textit{signature} $\leftarrow$ GetSignature()\;
        WriteFile(hFile: \textit{File}, lpBuffer: \textit{signature})\; \tcp{The signature is also written at the end}
\end{algorithm}

Therefore, the process is summarized as:

\begin{enumerate}
    \item Calculate the size of the buffer needed to hold an encrypted block of 0x2000 (8192) bytes.
    \item Obtain the size of the file.
    \item Encrypt the first 0x100000 bytes of the file in blocks of 0x2000 (8192) bytes.
    \item Write the victim ID (512 bytes) and the signature (24 bytes) at the end of the file.
\end{enumerate}

Here, we show an example of a signature written at the end of an encrypted file and highlight its different fields:\\

\colorbox{orange}{4E 4D 00 00 00 00 00 00} 00 02 00 00 01 00 00 00 \colorbox{cyan}{07 03 03 01} 01 01 E2 02\\

First, in orange, the original length of the file is written (0x4e4d or 20045 bytes in this case). Then, a hard-coded magic number is written at offset 16 (cian). This value is checked prior to encrypting a file, as we discussed earlier in this section.

\section{Decryption of infected systems} \label{ch:key_recovery}

In Section \ref{subsec:key_generation}, we described the functions responsible for importing, generating and destroying the cryptographic keys needed by the ransomware. As we pointed out, the key used for encrypting the system was randomly generated. Additionally, it was encrypted using a public key before being exported. Therefore, we are not able to know the key that is generated beforehand or to decrypt it after it has been exported, since we do not have the associated private key needed. However, we also hinted that the function responsible for destroying the cryptographic material was in fact never called if the system was not powered off. This is due to the ransomware process remaining in the background in order to encrypt new files or drives as they are created or connected. Since the keys are not destroyed and the ransomware process does not exit, we are able to recover the generated key. The only requirement is the memory of the ransomware process (i.e., a full dump). If such dump of the process (or the whole system) has been obtained, we may recover the key. This is of paramount importance, since users, upon seeing a ransom note, might be tempted to power off or reboot their systems in order to reestablish their machines, and would lose the opportunity of obtaining the key and thus decrypting the files.

In order to recover the key, we leverage the knowledge acquired during the advanced analysis process (see Section \ref{sec:analysis}) to identify the structure that points to the desired key. When a key is generated by using the Windows cryptography API (i.e., cryptsp.dll and rsaenh.dll) the key is an object of type HCRYPTKEY, which has the following structure \cite{CodeGuru2020}:

\noindent struct HCRYPTKEY\\
 \{ \\
\indent void* CPGenKey; \\
\indent void* CPDeriveKey; \\
\indent void* CPDestroyKey; \\
\indent void* CPSetKeyParam; \\
\indent void* CPGetKeyParam; \\
\indent void* CPExportKey; \\
\indent void* CPImportKey; \\  
\indent void* CPEncrypt; \\
\indent void* CPDecrypt; \\
\indent void* CPDuplicateKey; \\
\indent HCRYPTPROV hCryptProv; \\
\indent magic\_s *magic; \\
\}; \\

The first 10 fields of the structure point to functions of the Windows API. The eleventh field, \textit{hCryptProv}, points to the provider of the key and the functions (this provider must be first acquired before the key is generated via \textit{CryptAcquireContext} or a similar function). Finally, the last field points to another structure. This pointer is XOR-ed with a constant value, 0xE35A172C. Therefore, after XOR-ing the pointer with that magic constant, it points to the following structure:

\noindent struct magic\_s \\ 
\{ \\
\indent key\_data\_s *key\_data; \\
\}; \\

which contains a pointer to the following structure:

\noindent struct key\_data\_s \\
\{ \\
\indent void *unknown; \\
\indent uint32\_t alg; \\
\indent uint32\_t flags; \\
\indent uint32\_t key\_size; \\ 
\indent void* key\_bytes; \\
\}; \\

The \textit{key\_data\_s} structure contains three fields whose values are known:

\begin{itemize}
    \item \textit{alg} contains the algorithm ID of the algorithm for which the key has been generated. In this case, the value of this field is 0x00006610, which corresponds to AES256 \cite{MSAlgID}.
    \item \textit{flags} contains the value of the \textit{flags} parameter passed in the call to \textit{CryptGenKey} at 0x48f024. Therefore, its value is 0x00000001.
    \item \textit{key\_size}, as it name hints, contains the size of the key. In this case, the key is 32 bytes long (0x00000020).
\end{itemize}

Finally, the fifth field contains a pointer to the actual key. Since we know the value of 24 of the last 28 bytes that form the structure (skipping the first field) we can search for this 28-byte pattern in the memory of the process. We thus are able to obtain a pointer to the generated key that was used to encrypt the files and finally the key itself. 
We recall that the only requisite is that the system has not been powered off since it was infected, in order to maintain the key in memory. 

Now that we have the symmetric key generated by the ransomware, we are able to decrypt the infected files. However, to do so we need to implement the reverse operation than the one performed by the ransomware (see Algorithm \ref{alg:file_encryption}). To decrypt any given file, we first parse the signature at the end of the file. There, we obtain the original size of the encrypted file. Then, we truncate the file to eliminate both the signature and the block of 512 bytes appended at the end of the file by the ransomware (536 bytes in total, since the signature is 24 bytes in length). Once we have the truncated file, we proceed to decrypt the first 0x100000 bytes in blocks of 8192 (0x2000) bytes. Notice that, as we showed in Algorithm \ref{alg:file_encryption}, the \textit{Final} parameter in the \textit{CryptEncrypt} calls was never set to \textit{True}. According to the documentation, this parameter should be \textit{True} when the last block is encrypted. Although we do not know if this nonstandard behavior is intentional or not, we are forced to do the same in the decryption routine. Therefore, we always set the Final parameter to be False in the calls to \textit{CryptDecrypt}. Then, we copy the rest of the file as is. Finally, if the file was smaller than 0x100000 bytes, we truncate it once again, now to the original size recovered earlier from the signature appended at the end, to remove the padding bytes.

Obtaining a memory dump of a process can be done by standard forensic tools. Therefore, we open source the tool to recover the symmetric key from memory and decrypt the infected files:\\ \url{https://github.com/JavierYuste/AvaddonDecryptor}.

\section{Experimentation} \label{sec:results}

We test our proposal in a virtual environment running a Windows 7 x64 OS. In particular, we build this virtual machine on top of a virtualization solution named VirtualBox in a 1.60 GHz Intel Core i5-8250U CPU with 16 GB RAM computer. From the available hardware, we assign 2 cores and 4 GB of RAM to the aforementioned guest system.

Then, we execute Avaddon on the virtual machine and let it encrypt the whole system. When Avaddon has not utilized more than 0.5\% of the CPU time in the last 60 seconds, we understand that it has finished encrypting files and confirm the infection due to the presence of ransom notes and encrypted files through the whole file system.

After infecting the virtual machine, we proceed to decrypt all the affected files. First, we pause the ransomware process with Process Explorer, a tool from the SysInternals suite~\footnote{\url{https://docs.microsoft.com/en-us/sysinternals/}}. Note that we can freely drop executable files in the system before stopping Avaddon, since the \textit{exe} extension is excluded. Once the process is suspended, we can safely operate in the infected system. Next, we dump the memory of the ransomware process with ProcDump, which is also part of the SysInternals suite. Finally, we execute the proposed decryption tool, which we open source. This tool i) confirms the infection by extracting the signature appended at the end of encrypted files, ii) obtains the AES256 symmetric key from the dumped memory of the ransomware process iii) and decrypts the whole file system.

We show the results in Table \ref{tab:experimentation_results}. From 209,186 files that were present in the whole system, we found that 9,135 (4.3\%) were encrypted, making a total of 607 MB. Our proposed tool successfully decrypted all the affected files in 10 minutes and 35 seconds. Additionally, we have tested our tool with the most recent version of Avaddon, which was observed from a wild URL on mid-January 2021, when this paper was written. We confirm that the decryptor still works, since we were able to decrypt all the infected files. 

We must note some considerations. First, it is important to not turn off the computer after infection, since the proposed approach needs the encryption key to be present in memory. Otherwise, this would be destroyed and could only be recovered by means of the official channel proposed by the criminals, i.e. paying the ransom. Second, the proposed tool needs the original version of at least one encrypted file to find the correct symmetric key. This, however, can be easily achieved, e.g. by obtaining known files present by default in the Windows OS version installed in the affected system.

\begin{table}[]
    \centering
         \begin{tabular}{|c|c|} 
         \hline
         Files in the system & 209,186\\
         \hline
         Files encrypted by Avaddon & 9,135\\ 
         \hline
         Total size of files in the system & 46.85 GB\\
         \hline
         Total size of encrypted files & 607 MB\\
         \hline
         Time spent decrypting files & 558.54 s\\
         \hline
         Total time & 635.63 s\\
         \hline
        \end{tabular}
    \caption{Results of the experimentation in a virtual environment.}
    \label{tab:experimentation_results}
\end{table}

\section{Conclusions} \label{ch:conclusions}

Current approaches of cybercrime specialization, including new malware techniques, increase the threat of modern ransomware campaigns. In this work, we have analyzed a new ransomware, Avaddon, operated as a RaaS in a shared profit scheme, first seen on June 2020. Avaddon incorporates two techniques aimed at increasing their financial revenues which are growing in popularity: i) threatening victims that do not want to pay the ransom fee to leak personal data from infected systems, and ii) conducting DDoS attacks against them. Data leakage have affected at least 23 organizations whose information is allegedly exposed online. 
While having proper attribution is difficult, our analysis suggests that the threat actor behind Avaddon is from a CIS country. Indeed the initial announcement of the ransomware was made in a Russian underground forum, and it implements a policy to prevent infection of CIS-based victims. Moreover, a typographic error found in one of the processes fingerprinted by Avaddon suggest that this family is related with a previous ransomware, i.e. MedusaLocker, where the same error is also present. Indeed, the \textit{modus operandi} of Avaddon, that we detailed in this work, is similar to that of MedusaLocker \cite{TAU2020} and the list of services to stop is almost identical in both cases. 

By examining a sample obtained from the first campaign of Avaddon and describing its behaviors, we took a grasp on the general ``Cyber Kill Chain'' of ransomware threats (land, escalate privileges, deactivate defenses, acquire persistence, delete backups and encrypt files) and a detailed analysis of this ransomware in particular. Using an hybrid scheme, Avaddon attempts to hide the session key from defenders. However, due to the way in which cryptography keys are managed in this ransomware, we have developed a tool to recover the session key from the memory of the infected systems and decrypt all the affected files. The decryption tool also works with newer variants of the ransomware. The only requirement for this method to work is that the victim's computer is not powered off after the infection. 

Due to novelty of the ransomware, the business model in terms of an affiliate program, and the ability to extortion and blackmail victims (by means of exfiltration and DDoS attacks), it is likely to expect new variants of Avaddon and similar ransomware samples improving their mechanisms and expanding in the future. Thus, we believe that the analysis and tools provided in this paper can contribute to guide future analyses of such variants and to improve existing mitigation mechanisms.

\section*{Acknowledgements}

This work was supported by the Comunidad de Madrid (P2018/TCS-4566, co-financed by European Structural Funds ESF and FEDER).

\section*{References}

\bibliography{mybibfile}

\newpage
\section*{Appendix A} \label{appedix:list_whitelisted_strings}

\begin{table}[h!]
    \centering
         \begin{tabular}{|l|} 
         \hline
         ``C:\textbackslash Program Files\textbackslash Microsoft\textbackslash Exchange Server''\\
         \hline
         ``C:\textbackslash Program Files (x86)\textbackslash Microsoft\textbackslash Exchange Server''\\ 
         \hline
         ``C:\textbackslash Program Files\textbackslash Microsoft SQL Server''\\
         \hline
         ``C:\textbackslash Program Files (x86)\textbackslash Microsoft SQL Server''\\
         \hline
         ``C:\textbackslash Windows''\\
         \hline
         ``C:\textbackslash Program Files''\\
         \hline
         ``C:\textbackslash Users\textbackslash All Users''\\
         \hline
         ``C:\textbackslash Users\textbackslash Public''\\
         \hline
         ``C:\textbackslash Users\textbackslash \%User Profile\%\textbackslash AppData\textbackslash Local\textbackslash Temp''\\
         \hline
         ``C:\textbackslash Program Files (x86)''\\
         \hline
         ``C:\textbackslash Users\textbackslash \%User Profile\%\textbackslash AppData''\\
         \hline
         ``C:\textbackslash ProgramData''\\
         \hline
         ``Tor Browser''\\
         \hline
         ``AppData''\\
         \hline
         ``ProgramData''\\
         \hline    
         ``Program Files''\\
         \hline
         ``Windows''\\
         \hline
         Name of the ransom note (e.g., ``363053-readme.html'')\\
         \hline
         ``bckgrd.bmp''\\
         \hline
        \end{tabular}
    \caption{List of whitelisted strings in the encryption process.}
    \label{tab:list_whitelisted_strings}
\end{table}

\end{document}